\documentclass[10pt,conference]{IEEEtran}
\usepackage{pifont}
\usepackage{bbding}
\usepackage{amsmath}
\usepackage{amssymb}
\IEEEoverridecommandlockouts
\def\BibTeX{{\rm B\kern-.05em{\sc i\kern-.025em b}\kern-.08em
    T\kern-.1667em\lower.7ex\hbox{E}\kern-.125emX}}
\usepackage{graphicx}
\usepackage{amsthm}
\usepackage{subfigure,balance}
\usepackage{bm}
\usepackage{cite}

\usepackage{algorithm}% http://ctan.org/pkg/algorithms
\usepackage{amssymb, amsmath, amsfonts, epsfig, latexsym, times,algorithmicx}
\usepackage{balance,setspace}

\begin{document}
\title{Cache-enabled Wireless Networks with Opportunistic Interference Alignment}

\author{\IEEEauthorblockN{Y. He and S. Hu}
\IEEEauthorblockA{Department of Systems and Computer Eng., Carleton University, Ottawa, ON, Canada}
}
\maketitle

\begin{abstract}
Both caching and interference alignment (IA) are promising techniques for future wireless networks. Nevertheless, most of existing works on cache-enabled IA wireless networks assume that the channel is invariant, which is unrealistic considering the time-varying nature of practical wireless environments. In this paper, we consider realistic time-varying channels. Specifically, the channel is formulated as a finite-state Markov channel (FSMC). The complexity of the system is very high when we consider realistic FSMC models. Therefore, we propose a novel big data reinforcement learning approach in this paper. Deep reinforcement learning is an advanced reinforcement learning algorithm that uses  deep $Q$ network  to approximate the $Q$ value-action function.  Deep reinforcement learning is used in this paper to obtain the optimal IA user selection policy in cache-enabled opportunistic IA wireless networks. Simulation results are presented to show the effectiveness of the proposed scheme.

\end{abstract}


\section{Introduction}

Recently, \emph{information-centric networking} (ICN) has attracted great attentions from both academia and industry \cite{LYZ15}. In ICN, \emph{in-network caching} can efficiently reduce the duplicate content transmissions in networks. Caching has been recognized as one of the promising techniques for future wireless networks to improve spectral efficiency, shorten latency, and reduce energy consumption \cite{FYH15,liu2016caching}. %Proactively caching the popular contents can remove the heavy burden of the backhaul links. 

Another new technology called \emph{interference alignment} (IA) has been studied extensively as a revolutionary technique to tackle the interference issue in wireless networks \cite{Cadambe08,ZYJ16}. IA exploits the cooperation of transmitters to design the precoding matrices, and thus eliminating the interferences. IA can benefit mobile cellular networks \cite{suh2008interference}. Due to the large number of users in cellular networks, multiuser diversity has been studied in conjunction with IA, called opportunistic IA, which further improves the network performance \cite{perlaza2010spectrum, ZYL15_a,jung2011opportunistic, he2016multiuser, perlaza2008opportunistic}. 

Jointly considering these two important technologies, caching and IA, can be beneficial in IA-based wireless networks \cite{deghel2015benefits,maddah2015cache,ZLY16}. The implementation of IA requires the channel state information (CSI) exchange among transmitters, which usually relies on the backhaul link. The limited capacity of backhaul link has significant impacts on the performance of IA \cite{Ayach2013}. Caching can relieve the traffic loads of backhaul links, thus the saved capacity can be used for CSI exchange in IA.  In \cite{deghel2015benefits}, the authors  investigate the benefits of caching and IA in the context of mutiple-input and multiple-output (MIMO) interference channels, and maximize the average transmission rate by optimizing the number of the active transceiver pairs. In \cite{maddah2015cache}, it is shown that by properly placing the content in the transmitters' caches, the IA gain can be increased.

Although some excellent works have been done on caching and IA, most of these previous works assume that the channel is block-fading channel or invariant channel, where the estimated CSI of the current time instant is simply taken as the predicted CSI for the next time instant. Considering the time-varying nature of wireless environments, this kind of memoryless channel assumption is not realistic\cite{yang2005statistical,LYH10}. In addition, it is difficult to obtain the perfect CSI due to channel estimation errors, communication latency, handover and backhaul link constraints \cite{XYJ12,MYL04,YK07,BYY15}.

%On the other hand,  Cadambe and Jafar initially proposed IA and proved that only under the condition of varying channel coefficients can the $K$-user (without the limitation of $K\leqslant{3}$ ) interference channel achieve totally $KM/2$ DoFs with $M$ antennas at each node \cite{Cadambe08}.  Therefore, 

In this paper, we consider realistic time-varying channels, and propose a novel \emph{big data deep reinforcement learning} approach in cache-enabled opportunistic IA wireless networks. Cache-enabled opportunistic IA is studied under the condition of time-varying channel coefficients. The channel is formulated as a finite-state Markov channel (FSMC) \cite{wei2010distributed}. The complexity of the system is very high when we consider realistic FSMC models. Therefore, we propose a novel big data reinforcement learning approach in this paper. Deep reinforcement learning is an advanced reinforcement learning algorithm that uses  deep $Q$ network  to approximate the $Q$ value-action function \cite{mnih2015human}. Google Deepmind adopts this method on some games \cite{mnih2015human,SHMGS16}, and gets quite good results. Deep reinforcement learning is used in this paper to obtain the optimal IA user selection policy in cache-enabled opportunistic IA wireless networks. Simulation results are presented to illustrate that the performance of cache-enabled opportunistic IA networks can be significantly improved by using the proposed big data reinforcement learning approach.

%The realization of IA is based on the assumption that the global CSI is perfectly known at all active transmitters. This means that the CSI estimation and CSI exchange among active transmitters are two key issues. In \cite{xu2014time}, the authors show some practical challenges in obtaining the perfect CSI and provide schemes of time-invariant channel prediction for IA.  On the other hand, some codebook-based quantization techniques for compressing the channel matrices are used in the process of CSI exchange for reducing the backhaul link capacity\cite{deghel2015benefits}.  In this paper, we assume that the local CSI can be obtained perfectly, and fixed  backhaul link capacity $C_c$ is used for CSI exchange for each active transmitter. 

%Jointly consider the benefits of caching, opportunistic IA and time-varying channels, in this paper we investigate the optimization of cache-enabled opportunistic IA network with time-varying channel coefficients. This problem can be formulated as a stochastic selection process, and the recent advances of deep reinforcement learning can be utilized to find the optimal policy\cite{mnih2015human}. The distinct features of the paper are as follows.

The rest of this paper is organized as follows. Section II presents the system model. The deep reinforcement learning algorithm is presented in Section III. In Section IV, the system is formulated. Simulation results are discussed in Section V. Finally, Section VI gives the conclusions.

\section{System Model}
In this section, we describe the model of IA, followed by the time-varying channel. Then, cache-equipped transmitters are described.

\subsection{Interference Alignment}
\begin{figure}[t]
\begin{center}
\includegraphics[width=0.45\textwidth]{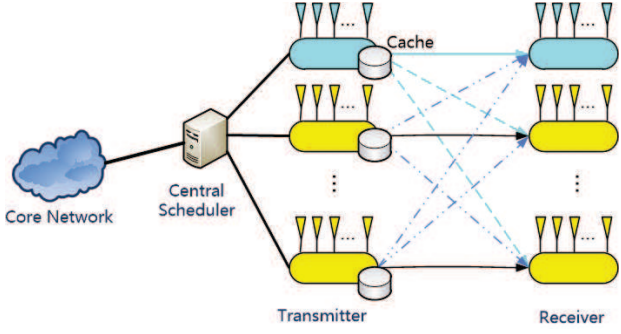}
\caption{System model of a $L$-user cache-enabled opportunistic IA wireless network.}\label{ia}
\end{center}
\end{figure}

%\label{sect:}
We consider a $L$-user MIMO interference network with limited backhaul capacity and caches equipped at the transmitter side, as illustrated in Fig. \ref{ia}.  There is a central scheduler who is responsible for collecting the channel state and cache status from each user, scheduling the users and allocating the limited resources. All the  users are connected to the central scheduler via a backhaul link for CSI share and Internet connection, and the total capacity is limited.

IA is a revolutionary interference management technique, which theoretically enables the network's sum rate grow linearly with the cooperative transmitter and receiver pairs. That is to say, each user can obtain the capacity $\frac{1}{2} \mbox {log}(\mbox{SNR})+o(\mbox {log}(\mbox {SNR}))$, which has nothing to do with the interferences.

Consider a $K$-user MIMO interference channel. $N_t^{[k]}$ and $N_r^{[k]}$ antennas are equipped at the $k$th transmitter and receiver, respectively. The number of data streams of the $k$th user is denoted as  $d^{[k]}$. The received signal at the $k$th receiver can be written as
\begin{equation}
\begin{aligned}
\textbf{y}^{[k]}\!(t)\!&=\!\textbf{U}^{[k]\dagger}(t)\textbf{H}^{[kk]}(t)\textbf{V}^{[k]}(t)\textbf{x}^{[k]}\!(t)\\
&+\!\!\!\!\!\!\sum\limits_{j = 1,j\neq
k}^K\!\!\!\!\textbf{U}^{[k]\dagger}(t)\textbf{H}^{[kj]}(t)\textbf{V}^{[j]}(t)\textbf{x}^{[j]}\!(t)\!+\!\textbf{U}^{[k]\dagger}\!(t)\textbf{z}^{[k]}\!(t),
\end{aligned}
\end{equation}
where the first term at the right side represents the expected signal, and the other two terms mean the inter-user interference and noise, respectively.
$\textbf{H}^{[kj]}(t)$ is the
$N_r^{[k]}\times{N_t^{[j]}}$ matrix of channel coefficients from the
$j$th transmitter to the $k$th receiver over the time slot $t$. Each element of $\textbf{H}^{[kj]}(t)$ is independent and identically distributed (i.i.d) complex Gaussian random variable, with zero mean and unit variance.
$\textbf{V}^{[k]}(t)$ and $\textbf{U}^{[k]}(t)$ are
the unitary $N_t^{[k]}\times{d^{[k]}}$ precoding matrix and
$N_r^{[k]}\times{d^{[k]}}$ interference suppression matrix of the \emph{k}th user, respectively. $\textbf{x}^{[k]}(t)$ and $\textbf{z}^{[k]}(t)$ are the
transmitted signal vector of $d^{[k]}$ DoFs and the $N^{[k]}\times{1}$ additive white Gaussian noise (AWGN) vector whose elements have zero mean and $\sigma^2$ variance at the $k$th receiver, respectively. 

The interference can be perfectly eliminated  only when the following conditions can be satisfied
\begin{equation}
\textbf{U}^{[k]\dagger}(t)\textbf{H}^{[kj]}(t)\textbf{V}^{[j]}(t)=0,\hspace{1mm}\forall
j\neq k,
\end{equation}
\begin{equation}
\mathrm{rank}\left(\textbf{U}^{[k]\dagger}(t)\textbf{H}^{[kk]}(t)\textbf{V}^{[k]}(t)\right)=d^{[k]}.
\end{equation}
Under this assumption, the received signal at the $k$th receiver can be rewritten as
\begin{equation}
\textbf{y}^{[k]}(t)=\textbf{U}^{[k]\dagger}(t)\textbf{H}^{[kk]}(t)\textbf{V}^{[k]}(t)\textbf{x}^{[k]}\!(t)+\textbf{U}^{[k]\dagger}(t)\textbf{z}^{[k]}(t).
\end{equation}

%The perfect implementation of IA is based on some strict assumptions\cite{Cadambe08}. 

To meet Condition (3), the global CSI is required at each transmitter. Each transmitter can estimate its local CSI (i.e., the direct link), but the  CSI of other links can only be obtained by CSI share with other transmitters via the backhaul link\cite{deghel2015benefits}. Thus, in IA network, the backhaul link is more than a pipeline for connecting with Internet. The limited capacity should be made optimum use of.  The recent advances focus on the benefits of edge caching, which is capable to decrease the data tranfer and leave more capacity for CSI share. The detail is described in the following subsection. In this paper, we assume the total backhaul link capacity of all the users is $C_{total}$, and the CSI estimation is perfect with no errors and no time delay.

\subsection{Time-varying Channel}
%\label{sect:}
We consider realistic time-varying channels in this paper. Since finite-state Markov channel (FSMC) is an effective model to characterize the fading nature of wireless channels\cite{wei2010distributed}, we choose FSMC model in this paper. Specifically, the first-order FSMC is used in this paper.
 
The received SNR is a proper parameter that can be used to reflect the quality of a channel. We model SNR  as a random variable, partition and quantize the range of the SNR into $H$ Levels, which is characterized by a set of states $\Upsilon=\{\Upsilon_0,\Upsilon_1, \Upsilon_2,\ldots, \Upsilon_{H-1}\}$. 
 We consider $T$ time slots over a period of wireless communication.  Let's denote $t\in\{{0,1,2,\ldots,T-1}\}$ as the time instant, and the SNR varies from one state to another state when one time slot elapses. 
 
Actually, SNR plays a crucial role in determining the IA results.  Cadambe and Jafar pointed out that IA performs better at very high SNR, and suffers from low quality at moderate SNR levels.  Meanwhile higher and higher SNR is required to approach IA network's theoretical maximum sumrate as the number of IA users increases\cite{Cadambe08}.  Thus, there exist competitions among users for accessing to IA network. 

\subsection{Cache-equipped Transmitters}
%\label{sect:cache-e}
 
In the era of explosive information, the vast amount of content makes it impossible for all of them gain popularity. As a matter of fact, only a small fraction becomes extensively popular.
That means certain content may be requested over and over during a short time span, which gives rise to the network congestion and transmission delay. We assume that each transmitter is equipped with a cache unit that has certain amount of storage space. The stored content may follow a certain popularity distribution. 

For consistency, the cache of each transmitter stores the same content,  usually the web content, and thus alleviating the backhaul burden and shorten delay time.  In \cite{tatar2014survey}, the authors survey on the existing methods for predicting the popularity of different types of web content. Specifically, they show that different types of content follow different popularity distributions. For example, the popularity growth of online videos complies with power-law or exponential distributions, that of the online news can be represented by power-law or log-normal distributions, etc. Based on the content popularity distribution and cache size, cache hit probability $P_{hit}$ and cache miss probability $P_{miss}$ can be derived\cite{deghel2015benefits}. In this paper, the specific popularity distribution is not the focus, and we just concentrate on two states, whether the requested content is within the cache or not. We describe the two states as $\Lambda=\{0,1\}$, where $0$ means the requested content is not within the cache, and $1$ indicates it is within the cache. %The probabilities $P_{hit}$  and $P_{miss}$ can be seen as the transitional probabilities between the two states. 

\section{Deep Reinforcement Learning}
In this section, we first present reinforcement learning. Then, deep Q-learning is described. 
\subsection{Reinforcement Learning}
%\label{sect:}
Reinforcement learning is an important branch of machine learning, where an agent makes interactions with an environment trying to control the environment to its optimal states that receive the maximal rewards.  The task of reinforcement learning can usually be described as a Markov Decision Process (MDP), however, state space, explicit transition probability and reward function are not necessarily required\cite{ong2015distributed}. Therefore, reinforcement learning is promising in handling tough situations that approach real-world complexity\cite{mnih2015human}.  

%Reinforcement learning can be classified into model-free and model-based reinforcement learning, which is based on whether the transition probability is given or not. Moreover, a variety of model-free and model-based algorithms can be used to approximate the reward functions\cite{gu2016continuous}. Let's briefly review the reinforcement learning from model-based to model-free case.  

Let $X = \{x_{1}, x_{2}, ...,x_{n}\}$ be  the state space, and
$A = \{a _{1}, a _{2}, ...a_{m}\} $ be the action set.
Based on the current state $x(t) \in X$, the  agent takes an action $a(t)\in A$
on the environment and then
the system transfers to a new state $x(t+1)\in X$ according to the transition probability $P_{x(t) x(t+1)}(a)$. The immediate reward is denoted as $r(x(t),a(t))$.

Taking into the long-term returns, the agent should not only consider the immediate rewards, but also the future rewards. The more into the future, the more discounts the reward may get. Thus, the future rewards are discounted with a discount factor $0<\epsilon<1$. 
The aim of the reinforcement learning agent is to find an optimal policy
$a^{*}=\pi^{*}(x)\in A$ for each state $x$, which maximizes the cumulative reward over a long time. The cumulative
discounted reward at state $x$ can be expressed by the state value function:
\begin{equation}
%\label{Vfunc}
V^{\pi}(x) = E\left[ \sum_{t=0}^{\infty}\epsilon^{t}r(x(t),a(t))|x(0)=x\right],
\end{equation}
where  $E$ denotes the
expectation, and it is considered over an infinite time horizon.

Due to the Markov property, i.e., the state at the subsequent time instant is only determined by the current state, irrelevant to the former states,  the value function can be rewritten as
\begin{equation}
%\label{Vjfunc1}
V^{\pi}(x) = R(x,\pi(x))+\epsilon\sum_{x{'}\in X}P_{x x^{'}}(\pi(x))V^{\pi}(x^{'}),
\end{equation}
where  $R(x,\pi(x))$ is the mean value of the immediate reward $r(x,\pi(x))$, and $P_{x x^{'}}(\pi(x))$ is the transition probability from $x$ to $x^{'}$, when action $\pi(x)$ is executed.
The optimal policy $\pi^{*}$ follows Bellman's criterion
\begin{equation}
%\label{Vjfunc1}
V^{\pi^{*}}(x) = \max_{a^{'}\in A}\left[R(x,a)+\epsilon\sum_{x{'}\in X}P_{x x^{'}}(a)V^{\pi^{*}}(x^{'})\right].
\end{equation}

Given the reward $R$ and transition probability $P$, the optimal policy can be obtained.

\subsection{Deep Q-learning}
%\label{sect:}
When $R$ and $P$ are unknown, Q-learning is one of the most widely-used strategies to determine the best policy ${\pi^{*}}$. A state-action function, i.e., Q-function is defined as
\begin{equation}
%\label{Vjfunc1}
Q^{\pi}(x,a) = R(x,a)+\epsilon\sum_{x{'}\in X}P_{x x^{'}}(a)V^{\pi}(x^{'}),
\end{equation}
which represents the discounted cumulative reward when action $a$ is performed at state $x$ and continues optimal policy from that point on.

The maximum Q-function will be
\begin{equation}
%\label{Vjfunc1}
Q^{\pi^{*}}(x,a) = R(x,a)+\epsilon\sum_{x{'}\in X}P_{x x^{'}}(a)V^{\pi^{*}}(x^{'}),
\end{equation}
then the discounted cumulative state function can be written as
\begin{equation}
%\label{Vjfunc1}
V^{\pi^{*}}(x) =\max_{a\in A}\left[Q^{\pi^{*}}(x,a)\right].
\end{equation}

Up to now, the objective can change from finding the best policy to finding the proper Q-function. Usually, Q-function is obtained in a recursive manner using the available information $(x, a, r, x^{'}, a^{'})$, i.e., the state $x$, the immediate reward $r$, the action $a$ at the current time instant $t$, and the state $x^{'}$ and action $a^{'}$ at the next time instant $t+1$. The Q-function is  updated as
\begin{align}
%\label{Vjfunc1}
Q_{t+1}(x,a)  =  & Q_t(x,a) +  \nonumber \\
& \alpha\!\left(\!r+\epsilon [{\max_{a^{'}} Q_t(x^{'},a^{'})}]-Q_t(x,a)\right),
\end{align}
where $\alpha$ is the learning rate. Utilizing proper learning rate, $Q_t(x,a)$ will definitely converges to $Q^{*}(x,a)$\cite{nie1999q}. 

As a matter of fact, the Q-function is commonly estimated by a function approximator, sometimes a nonlinear approximator, such as a neural network $  Q(x,a;\theta) \approx  Q^{*}(x,a) $. This neural network is named $Q$ network. The parameter $\theta$ are the weights of the neural network, and the network is trained by adjusting $\theta$ at each iteration to reduce the mean-squared error. 

However,  $Q$-network exhibits some instabilities, and the causes are provided in \cite{mnih2015human}.
Deep $Q$ learning, in which deep neural network is used to approximate the Q-function,  is proposed recently, and it is proven to be more advantageous\cite{mnih2015human}.  Two techniques were used by deep Q-learning to modify the regular Q-learning. The first one is experience replay. At each time instant $t$, an agent stores its interaction experience tuple $e(t) =(x(t) ,a(t) ,r(t) ,x(t+1))$  into a replay memory $D(t)= \{e(1) ,...,e(t)\}$. Then it randomly samples from the experience pool to train the deep neural network's parameters rather than directly using the consecutive samples as in Q-learning.  The other modification is that deep Q-learning adjusts the target value to update several time steps, instead of updating every time step. The target value is expressed as $y=r+\epsilon \max_{a^{'}}Q(x^{'},a^{'},\theta_{i}^{-})$. 
In the Q-learning, the weights $\theta_{i}^{-}$ are updated as $\theta_{i}^{-}=\theta_{i-1}$, whereas in the deep Q-learning  $\theta_{i}^{-}=\theta_{i-N}$, i.e.,  the weights update every $N$ time steps. Such modification can make the learning process more stable. 

The deep $Q$ function is trained towards the target value by minimizing the loss function
$L(\theta)$ at each iteration, the loss function can be written as
\begin{equation}
%\label{Vjfunc1}
L(\theta)=E[(y-Q(x,a,\theta)^2)].
\end{equation}

We use deep reinforcement learning in optimizing the performance of the cache-enabled IA network, and the formulation process is described in the following section.

\section{Problem Formulation}
In this section, we formulate the cache-enabled IA network optimization problem as a deep $Q$-learning process, which can determine the optimal policy for IA user grouping.

In our system, there are $L$ candidates that want to join in the IA network to communicate wirelessly. We assume that the IA network size is always smaller than the number of candidates, which is in accordance with the fact that a large number of users expect wireless communications anytime and anywhere.  As aforementioned, the value of SNR affects the  performance of interference alignment, and the candidates who occupy the better channels are more advantageous for accessing to the IA network.  Therefore, we make an action at each time slot to decide which candidates are the optimal users for constructing an IA network based on their current states.

Here, a central scheduler is responsible for acquiring each candidate's CSI and cache status, then it assembles the collected information into a system state. Next, the controller sends the system state to the agent, i.e., the deep $Q$ network, and then the deep $Q$ network feeds back the optimal action $\arg \max_{\pi}Q^{*}(x,a)$ for the current time instant.
After obtaining the action, the central scheduler will send a bit to inform the users to be active or not, and the corresponding precoding vector will be sent to each active transmitter. 
The system will transfer to a new state after an action is performed, and the rewards can be obtained according to the reward function. 

Inside the deep $Q$ network, the replay memory stores the agent's experience of each time slot. The $Q$ network parameter $\theta$  is updated at every time instant with samples from the replay memory. The target $Q$ network parameter $\theta^{-}$ is copied from the $Q$ network every $N$ time instants. The $\varepsilon$-greedy policy is utilized to balance the exploration and exploitation, i.e., to balance the reward maximization based on the knowledge already known with trying new actions to obtain knowledge unknown. 
%The training algorithm of the deep $Q$ network is described in Algorithm \ref{alg}. 

%Actually, the optimal  policy can be learned off-line  after we identify the states, actions, reward functions in our system model.  The cost function is identified to calculate the off-line cost. Both the state transition probabilities and cost functions are identified for off-line simulations, and they will not be needed if we can carry out our deep $Q$ learning  model in real CBTC systems.

In order to obtain the optimal policy, it is necessary to identify  the actions, states and reward functions  in our deep $Q$ learning model, which will be described in the next following subsections.

%\begin{algorithm}[tb]
%\renewcommand{\thealgorithm}{{\em 1}}
%\caption{Deep reinforcement learning  algorithm in cache-enabled IA networks.}
%\label{alg}
%\begin{algorithmic}
 %\State  Initialization.
%\\
%\State \ \  Initialize  replay memory.
%
%\State \ \  Initialize the $Q$ network parameters with $\theta$.
%
%\State \ \  Initialize the target $Q$ network parameters with $\theta^{-}=\theta$.
%
%\ \ \State \textbf{For} $ \mbox{episode} \ \  k=1,...,K$ \textbf{do}:
%
%\\\State \ \  Initialize  the beginning state $x$.
%
%\\\State \ \  \textbf{For} $t= 1,2,3..., T$ \textbf{do}:
%\\
%\State \ \      \qquad  Choose a random probability $p$.
%\State \ \     \qquad   Choose $a(t)$ as,
%
%\qquad$\mbox {if }  p \geq \varepsilon$,\\
 %\qquad\qquad \qquad $a^{*}(t)=\arg \max_{a}Q(x,a,\theta), $\\
%\qquad\qquad $\mbox{otherwise}, $\\
%\qquad\qquad \qquad $\mbox{randomly\  choose\  a\  solution} \  a(t)\neq a^{*}(t)$,
%
%\State \ \  \qquad  Execute $a(t)$ in the system. Observe the reward $r(t)$, and next state $x(t+1)$. Store the experience $(x(t),a(t),r(t),x(t+1))$ in replay memory.
%
%\State \ \  \qquad  Get mini-batch of samples $(x(t),a(t),\zeta(t),x(t+1))$  from the replay memory.
%\State \ \  \qquad Perform a gradient descent step on $(y-Q(x,a,\theta)^2)$ with respect to $\theta$.\\
%\qquad Return the value parameters $\theta$ in Q network.
%
%\textbf{End for}\\
%\textbf{End for}
%
%
%
%\end{algorithmic}
%\end{algorithm}

\subsubsection{System State}
The current system state $x(t)$ is jointly determined by the states of $L$ candidates. The system state at time slot $t$ is defined as,
\begin{equation}
\label{State}
x(t) = \{\gamma_{1}(t),c_1(t),\gamma_{2}(t),c_2(t),\ldots, \gamma_{L}(t),c_L(t)\},
\end{equation}
where each candidate contains two states: the channel state $\gamma_{i}(t)\in{\Upsilon}=\{\Upsilon_0,\Upsilon_1,\ldots,\Upsilon_{H-1} \}$, and  the cache state $c_i(t)\in{\Lambda}=\{0,1 \}$, the index $i$ means the $i$th candidate, and $i=1,2,\ldots, L$.

The number of possible system states is ${(2\times H)}^L $, and this number can be very large as  $L$ increases. Due to the curse of dimensionality, it is difficult for traditional approaches handle our problem. Fortunately, deep $Q$ network is capable of successfully learning directly from high-dimensional inputs\cite{mnih2015human}, thus it is proper to be used in our system.

\subsubsection{System Action}
In the system, the central scheduler has to decide which candidates to be set active, and the corresponding resources will be allocated to the active users. 

The current composite action $a(t) $ is denoted by
\begin{equation}
\label{Action}
a(t) = \{a_{1}(t),a_{2}(t),\ldots , a_{L}(t) \},
\end{equation}
where $a_{i}(t)$ represents the control of the $i$th candidate, and 
each element $a_i{(t)}\in\{0,1\}$, and $a_i{(t)}=0$ means the candidate $i$ is passive (not selected) at time slot $t$, and $a_i{(t)}=1$ means it is active (selected).

\subsubsection{Reward Function}\label{cost}
Reward function indicates the received reward when a certain action is performed under a certain state.
The system reward represents the optimization objective, and we take the objective to maximize the IA network's throughput,  and the reward function of the $n$th candidate is defined as Eq. (\ref{refspeed}) on the top of the next page. Here, $C_{total}$ is the total capacity of the backhaul link, and $C_c$ is the fixed capacity allocated to each active user to exchange CSI with other active users. For the $n$th candidate, if the requested content is not in the local cache, it can only acquire the content from the backhaul link, and equal capacity (the total capacity minus the total capacity for CSI exchange) is allocated among the active users. If the requested content is within the cache, the $n$th candidate can get the maximum rate that an IA user can achieve. Note that, for simplicity we assume the interference can be perfectly eliminated, and each active user's sum rate is approaching half the capacity that the user could achieve without interferers.

%\qquad
\begin{figure*}
\begin{equation} \label{refspeed}
\begin{aligned}
    r_l(t) = 
	 \left\{
      \begin{array}{ll}
    {a_l(t)}\log_2 \left( 1+\displaystyle\frac{\left|\textbf{u}^{[l]\dagger}\textbf{H}^{[ll]}\textbf{v}^{[l]}\right|^2P^{[l]}x_l}{\sum\limits_{j=1,j\neq l}^{L}a_j(t)\left|\textbf{u}^{[l]\dagger}\textbf{H}^{[lj]}\textbf{v}^{[j]}\right|^2P^{[j]}x_j+\sigma^2}     \right), \mbox{if } c_l(t)=1, \\
      \\
     {a_l(t)}\min \left\{ \left[\displaystyle\frac{1}{\sum\limits_{i=1}^{i=L} a_i(t)}(C_{total}-C_c \sum\limits_{i=1}^{i=L} a_i(t))\right]\right., \left. \log_2 \left( 1+\displaystyle\frac{\left|\textbf{u}^{[l]\dagger}\textbf{H}^{[ll]}\textbf{v}^{[l]}\right|^2P^{[l]}x_l}{\sum\limits_{j=1,j\neq l}^{L}a_j(t)\left|\textbf{u}^{[l]\dagger}\textbf{H}^{[lj]}\textbf{v}^{[j]}\right|^2P^{[j]}x_j+\sigma^2}     \right)  \right\}\\ 
\\
    \qquad\qquad\qquad\qquad\mbox{if } c_l(t)=0,   \\
      \end{array}
    \right.
\end{aligned}
\end{equation}
\end{figure*}

The immediate system reward is the sum of all the candidates' immediate rewards, i.e., $r(t)=\sum\limits_{l=1}^{l=L}r_l(t) $.  
The central scheduler gets $r(t)$ in state $x(t)$ when action $a(t)$ is performed in time slot $t$. However, a maximum immediate value does not mean the maximum long-term cumulative rewards. Therefore, we should also think about the future rewards. The more into the future, the more uncertainty there exists. A discounted future reward with a discount factor $\epsilon$ is much more reasonable. The goal of using deep $Q$ network into our system model is to find a selection policy that maximizes the discounted cumulative rewards during the communication period $T$, and the cumulative reward can be expressed as 
\begin{equation}
%\label{Vjfunc1}
R = \max_{\pi} E\left[\sum\limits_{t=0}^{t=T-1}\epsilon^{t} r(t)\right] ,
\end{equation}
where $\epsilon^{t}$ approaches to zero when $t$ is large enough.  In practice, a threshold for terminating the process can be set.

\section{Simulation Results and Discussions}
In this section, computer simulations are carried out to demonstrate the performance of the proposed big data deep reinforcement learning approach to the optimization of cache-enabled opportunistic IA wireless networks. We compare the proposed scheme with two other schemes: 1) The same proposed approach without caching and 2) An existing user selection approach without cache \cite{ZYS16}, in which invariant channels are assumed. The performance improvements of the proposed scheme are present. 

In the simulations, we consider a cache-enabled opportunistic IA network, in which $L=5$ candidates want to access to. Due to the feasibility of IA\cite{yetis2010feasibility}, i.e., $N_t+N_r\geq{d(L+1)}$, we assume that each candidate is equipped with three antennas at both the transmitter node and the receiver node, and DoF is set to be 1. We quantize and partition the received SNR into 10 levels, i.e., [$-{\infty}$, 5], [5, 10], [10, 15], [15, 20], [20, 25], [25, 30], [30, 35], [35, 40], [40, 45] and [45, $+{\infty}$].  We assume that the channel state transition probability is identical for all the candidates. In one simulation scenario, the transition probability of remaining in the same state is set to be 0.489, and the probability of transition to the adjacent state to be twice that of transition to a nonadjacent state. %The channel state transition matrix is shown on the top of the next page. We change the channel state transition probability in other simulation scenarios. 
%\begin{figure*}
%\begin{eqnarray*}
%P_{channel} = \left(
   %\begin{array}{cccccccccc}
   %0.489 & 0.256 & 0.128 & 0.064  & 0.032 & 0.016 & 0.008 &0.004 & 0.002 & 0.001\\
   %0.001 & 0.489 & 0.256 & 0.128 & 0.064  & 0.032 & 0.016 & 0.008 &0.004 & 0.002\\
   %0.002&0.001 & 0.489 & 0.256 & 0.128 & 0.064  & 0.032 & 0.016 & 0.008 &0.004 \\
   %0.004&0.002&0.001 & 0.489 & 0.256 & 0.128 & 0.064  & 0.032 & 0.016 & 0.008 \\
   %0.008 &0.004 & 0.002&0.001 & 0.489 & 0.256 & 0.128 & 0.064  & 0.032 & 0.016 \\
   %0.016 & 0.008 &0.004 & 0.002 & 0.001& 0.489 & 0.256 & 0.128 & 0.064  & 0.032\\
   %0.032 & 0.016 & 0.008 &0.004 & 0.002&0.001 & 0.489 & 0.256 & 0.128 & 0.064 \\
   %0.064  & 0.032 & 0.016 & 0.008 &0.004& 0.002&0.001 & 0.489 & 0.256 & 0.128 \\
   %0.128 & 0.064  & 0.032 & 0.016 & 0.008 &0.004& 0.002&0.001 & 0.489 & 0.256 \\
   %0.256 & 0.128 & 0.064  & 0.032 & 0.016 & 0.008 &0.004& 0.002&0.001 & 0.489\\
   %\end{array}
 %\right).
%\end{eqnarray*}
%\end{figure*}

The cache at each transmitter includes two states: existence and nonexistence of the requested content. The implementation of the big data deep reinforcement learning algorithm is based on the TensorFlow to derive the optimal policy for IA user selection.   The discount factor $\epsilon$ is set to be 0.5, and the learning rate $\alpha$ is designed to be state-action dependent varying with time. In the $\varepsilon$-greedy exploration, $\varepsilon$ is initially set to be 0.1, and finally to be 1. The $Q$ value update frequency $N$ is set to be 4, and the relay memory size is 100K. 

%\begin{figure}[tb]
%\begin{center}
%\includegraphics[width=0.4\textwidth]{NNET.eps}
%\caption{The $Q$ network used in the deep reinforcement learning algorithm.}\label{q_network}
%\end{center}
%\end{figure}

\begin{figure}[tb]
\begin{center}
\includegraphics[bb=68 70 587 493,scale=0.44]{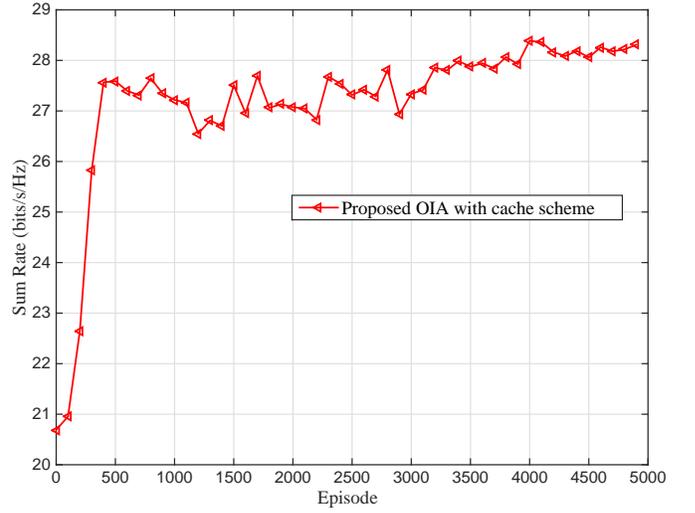}
\caption{Convergence performance of the proposed scheme.}\label{conv}
\end{center}
\end{figure}

\begin{figure}[tb]
\begin{center}
\includegraphics[bb=68 70 587 493,scale=0.44]{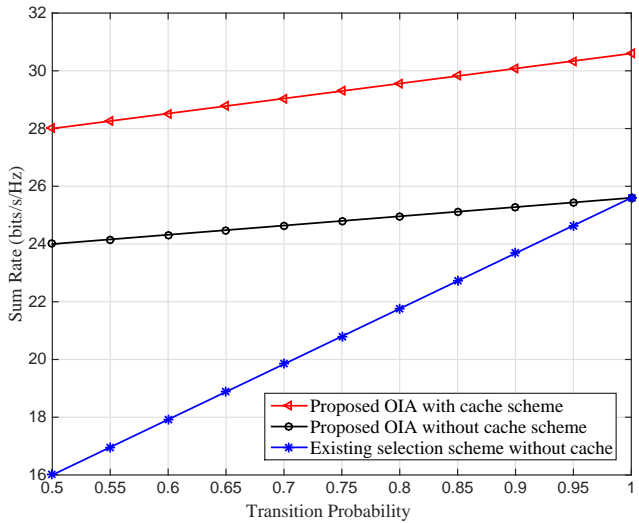}
\caption{Comparison of sum rate with different transition probabilities.}\label{prob}
\end{center}
\end{figure}

Fig. \ref{conv} shows the convergence performance of the proposed scheme. From this figure, we can  observe that the sum rate of the proposed scheme is low at the beginning of the learning process. During the learning process, the sum rate increases, and converges after about 3500 episodes. Please note that the learning is done off-line to train the deep neural network parameters. Fig. \ref{prob} shows the network's average sum rate with different state-transition probabilities of staying in the same state.  It can be seen that the proposed OIA with cache scheme can achieve the highest sum rate compared to the other two schemes. This is because the channel is time-varying, and the proposed scheme can obtain the optimal IA user selection policy in the realistic time-varying channel environment using the big data deep reinforcement learning algorithm. We can also observe that the performance of the existing selection method is getting closer to the proposed OIA without cache scheme as the transition probability increases, and this method performs the same when the channel remains absolutely static, i.e., the transition probability that the channel will be in the same state is 1. 

%[width=0.52\textwidth]

\section{Conclusions and Future Work}
In this paper, we studied cache-enabled opportunistic IA under the condition of time-varying channel coefficients. The system complexity is very high when we model the time-varying channel as a finite-state Markov channel. Thus, we exploited the recent advances, and formulated the system as a big data deep reinforcement learning problem. A central scheduler is responsible for collecting the CSI from each candidate, and then sends the integral system state to the deep $Q$ network to derive the optimal policy for user selection. Simulation results were presented to show that the performance of cache-enabled opportunistic IA networks can be significantly improved by using the proposed big data reinforcement learning approach. Future work is in progressed to consider wireless virtualization in the proposed framework.

\balance

\bibliography{reference}  
\end{document}